\documentclass[reprint,showkeys,onecolumn,
superscriptaddress,
 amsmath,
 amssymb,
 aps,
]{revtex4-2}

\usepackage[utf8]{inputenc}

\usepackage{graphicx}% Include figure files
\usepackage{dcolumn}% Align table columns on decimal point
\usepackage{bm}% bold math
\usepackage[colorlinks]{hyperref}
\hypersetup{
    linkcolor = blue,   % 超链接文本颜色
    citecolor = blue,      % 引用标记颜色
    urlcolor  = blue    % URL链接颜色
}
\usepackage{color}
\usepackage{amsmath}

\begin{document}

\title{Generally covariant geometric momentum and geometric potential for a Dirac fermion on a two-dimensional hypersurface}

\author{Zhao Li}
% \email{leezhao@hnu.edu.cn}
\affiliation{Basic Teaching Department, Nanhang Jincheng College, Nanjing 211156, China}

\author{Long-Quan Lai}
\email{lqlai@njupt.edu.cn}
\affiliation{School of Science, Nanjing University of Posts and Telecommunications, Nanjing 210023, China}

\date{\today}

\begin{abstract}
Geometric momentum is the appropriate momentum for a particle constrained to move on a curved surface, which depends on the extrinsic curvature and leads to observable effects, and curvature-induced quantum potentials appear for a nonrelativistic free particle on the surface. In the context of multi-component quantum states, the geometric momentum should be rewritten as a generally covariant geometric momentum, which contains an additional term defined as the gauge
 potential. For a Dirac fermion constrained on a two-dimensional hypersurface, we derive the generally covariant geometric momentum, and demonstrate that no curvature-induced geometric potentials arise on a pseudosphere or a helical surface. The dynamical quantization conditions are verified to be effective in dealing with constrained systems on hypersurfaces, enabling the derivation of both the generally convariant geometric momentum and the geometric potential for a spin particle constrained on parametrically defined surfaces.
\end{abstract}


\maketitle

\section{Introduction}

The quantum motion of particles constrained on a two-dimensional curved surface has become an increasingly relevant topic with the advancement of quantum technologies, ranging from quantum physics to condensed matter physics. Typical examples include the spectra of polyatomic molecules \cite{maraner}, electronic states on helicoidal and Beltrami surfaces \cite{jensen,atanasov,furtado}, curvature-induced modifications in the electronic properties of graphene \cite{belonenko,zhukov,champo}, and geometry-induced gauge fields \cite{brandt,schmidt,wang,mazharimousavi} and geometric potentials \cite{costa,ferrari,valle,szameit,onoe}. In classical mechanics, the Hamiltonian is expressed in terms of the intrinsic coordinates and momenta, and the motion of a nonrelativistic free particle on a surface solely depends on its intrinsic geometry. The classical equation of motion, $dF/dt=\left[F,H\right]
_{\rm{P}}$, where $\left[ \ldots \right]_{\rm{P}}$ denotes the Poisson bracket for an observable $F$, retains its form in quantum mechanics as $dF/dt=\left( 1/i\hbar \right) \left[F,H\right] $. This correspondence underpins the fundamental quantum conditions \cite{dirac} and arises naturally from the Dirac's canonical quantization, which allows direct construction of the quantum operators.

The fundamental quantum conditions refer to
a set of commutation relations between the components of coordinate and momentum operators, writing
\begin{eqnarray}
\left[ x_{i},x_{j}\right] =0,\text{ }
\left[ p_{i},p_{j}\right] =0,\text{ }
\left[ x_{i},p_{j}\right] =i\hbar \delta _{ij},\text{ }
\left(
i,j,k,l=1,2,3,\ldots, N\right).
\end{eqnarray}
Here, $x_{i}$ and $p_{j}$ denote, respectively, the coordinate operator and momentum
operator of a particle moving in $N$-dimensional Euclidean space $E^{N}$.
However, this framework no longer holds when the system is constrained. For a particle confined to a smooth curved surface $\Sigma ^{N-1}$ in $E^{N}$, the standard Poisson bracket $\left[ F,H\right]_{\rm{P}}$ must be replaced by the
Dirac bracket $\left[ F,H\right]_{\rm{D}}$ in canonical quantization procedures,
leading to the modified quantum conditions \cite{homma,klauder}
\begin{eqnarray}
\left[ x_{i},x_{j}\right] =0,\text{ }
\left[ x_{i},p_{j}\right] =i\hbar\left( \delta _{ij}-n_{i}n_{j}\right) ,\text{ }
\left[ p_{i},p_{j}\right]
=-i\hbar \left\{ \left( n_{i}n_{k},_{j}-n_{j}n_{k},_{i}\right) p_{k}\right\}
_{\rm{Hermition}},
\end{eqnarray}
where $O_{\rm{Hermition}}$ represents a properly constructed Hermitian
operator of an observable $O$. The hypersurface $\Sigma ^{N-1}$ can be
described by a constraint $f\left( {\bf{x}}\right)
=0$ in the configuration space and the equation of the surface is chosen as $\left\vert \nabla f\left(
{\bf{x}}\right) \right\vert =1$, such that the normal vector is given by ${\bf{n}}\equiv
\nabla f\left( {\bf{x}}\right) ={\bf{e}}_{i}n_{i}$. Consequently, only the unit normal vector and its derivatives appear, which is independent of the specific form of the surface equation \cite{li1,li2}.

Under the above quantum conditions, the quantum momentum ${\bf{p}}$ admits multiple possible forms due to the
operator-ordering ambiguity in $O\left\{ \left(
n_{i}n_{k},_{j}-n_{j}n_{k},_{i}\right) p_{k}\right\} _{\rm{Hermition}}$, which implies that even the
proper form of the momentum and the Hamiltonian cannot be uniquely determined without additional constraints \cite{ikegami,ogawa,ogawa1,ogawa2}. While the Hamiltonian is not the focus of the present work, we adopt the symmetry as a guiding principle in quantization procedures and aim to construct
additional commutation relations, such as $\left[{\bf{x}},H\right] =i\hbar\left[
{\bf{x}},H\right]_{\rm{D}}$ and $\left[ {\bf{p}},H\right] =i\hbar \left[
{\bf{p}},H\right] _{\rm{D}}$, allowing for the simultaneous quantization of the Hamiltonian, coordinates, and momenta, rather than presupposing specific operator forms. To go
beyond the operator-ordering problem, we have the dynamical quantization conditions as \cite{lian}
\begin{eqnarray}
\left[ {\bf{x}},H\right] &\equiv& i\hbar \frac{\mathbf{p}}{m_{0} }, \\
{\bf{n}}\wedge \left[ {\bf{p}},H\right] -\left[ {\bf{p}},H\right]
\wedge {\bf{n}} &=& 0.  \label{dqc}
\end{eqnarray}
Here, $m_{0}$ is the mass of the particle, and Eq.~(\ref{dqc})
indicates that the particle experiences no tangential force. The fundamental quantization conditions and the dynamical quantization conditions constitute the so-called enlarged canonical quantization
scheme, which gives the explicit form of the momentum as \cite{liu,liu1,liu2}
\begin{eqnarray}
{\bf{p}}=-i\hbar \left( \nabla _{\Sigma }+\frac{M{\bf{n}}}{2}\right) ,
\end{eqnarray}
where $\nabla _{\Sigma }\equiv {\bf{e}}_{i}\left( \delta _{ij}-n_{i}n_{j}\right) \partial
_{j}=\nabla _{N}-{\bf{n}}\partial _{n}$ is the gradient operator on the surface $\Sigma ^{N-1}$ with $\nabla _{N}$ being the standard gradient operator
in $E^{N}$. The quantity $M\equiv -\nabla _{N}\cdot {\bf{n}}=-\partial_{i}n_{i}$
denotes the mean curvature, in which repeated indices are summed over and that is in fact the trace of the extrinsic curvature tensor. We refer to ${\bf{p}}$ as the geometric momentum due to its dependence on the geometric
invariants. It satisfies the following simplest form of the commutation relation
\begin{eqnarray}
\left[ p_{i},p_{j}\right] =-i\hbar \left\{ \left(
n_{i}n_{k},_{j}-n_{j}n_{k},_{i}\right) p_{k}+p_{k}\left(
n_{i}n_{k},_{j}-n_{j}n_{k},_{i}\right) \right\} /2  
\end{eqnarray}
and the compatibility of constraint condition ${\bf{n}}\cdot {\bf{p}}+{\bf{p}}\cdot {\bf{n}}=0$, which means that in quantum mechanics the motion remains in the tangential
plane and corresponds to the constrained condition in classical mechanics: $
{\bf{n}}\cdot {\bf{p}}={\bf{p}}\cdot {\bf{n}}=0$ \cite{liu,liu1,liu2}. The geometric momentum, which arises purely from quantum considerations, depends
on the extrinsic geometry of the embedding of $\Sigma ^{N-1}$ in $E^{N}$. The extrinsic dependence highlights a fundamental distinction between classical and quantum confinements. Notably, the geometric momentum has been experimentally verified
\cite{spittel}, indicating that a purely intrinsic geometric framework is insufficient for accurately describing constrained quantum motions, unless the extrinsic examination is also performed.

The geometric momentum is sufficient for describing single-component quantum states. When dealing with multi-component quantum states \cite{lee,iorio}, however, the gauge structure should also be included. With the gradient operator $\nabla_{\Sigma}={\bf{e}}_{i}\left(\delta_{ij}-n_{i}n_{j}\right)={\bf{r}}^{\mu}\partial_{\mu}$ and a transformation $\partial_{\mu}\rightarrow D_{\mu}=\partial_{\mu}+i\Omega_{\mu}$, we immediately obtain the generally covariant geometric momentum \cite{liu3}
\begin{equation}
\mathbf{p}=\mathbf{-}i\hbar \left( \nabla _{\Sigma }+\frac{M\mathbf{n}}{2}%
\right) -\mathbf{A,}  
\end{equation}
where $\mathbf{A}=\hbar{\bf{r}}^{\mu}\Omega_{\mu}$ denotes the gauge potential, with $\Omega_{\mu}=\frac{i}{8}\omega_{\mu}^{ab}\left[\gamma_{a},\gamma_{b}\right]$, $\omega_{\mu}^{ab}$ being the spin connection, and $\gamma_{a}$ representing the Dirac spin matrix \cite{iorio}. One can rewrite the quantity $\Omega_{\mu}$ as the product of $\frac{1}{4}\omega_{\mu}^{ab}$ and $Q_{ab}=\frac{1}{2i}\left[\gamma_{a},\gamma_{b}\right]$, and take the eigenvalues of the matrix $Q_{ab}$ as the effective interaction strengths. The generally covariant geometric momentum in this form is applicable to both relativistic and nonrelativistic particles regardless of the mass.

The celebrated curvature-induced geometric potential has been experimentally confirmed \cite{szameit,onoe}, establishing its essential role in describing the particle motion on curved surfaces. For nonrelativistic particles, a basic framework for characterizing both geometric momentum and geometric potential has been established \cite{liu,liu1,liu2}, while the geometric momentum should be rewritten as a generally covariant geometric momentum in the context of multi-component quantum states \cite{liu3}.For a Dirac fermion constrained typically on a two-dimensional curved surface of revolution, such as the torus, catenoid and symmetric ellipsoid, there is no existence of the geometric potential \cite{yang}, and a natural question thus arises as to how this feature manifests in two-dimensional hypersurfaces. In the present work, we demonstrate the formalism of deriving the generally covariant geometric momentum and the geometric potential on a hypersurface, and show that for two-dimensional cases, e.g., the pseudosphere and the helical surface, the geometric potential takes the form of a constant matrix independent of surface parameters, which is composed of the $z$-direction Pauli matrix and the identity matrix. There is currently not a general result for a constrained Dirac fermion on arbitrary two-dimensional hypersurfaces. The clear framework, however, facilitates the access to both the generally covariant geometric momentum and the geometric potential for parametrically defined surfaces.

The rest of the paper is organized as follows. In Sec.~\ref{geometric}, we give the generally covariant geometric momentum and the geometric potential, respectively, for a Dirac fermion that is constrained on a curved surface with a formal parametric equation. In Sec.~\ref{example}, we present two typical examples with a two-dimensional pseudosphere and a helical surface as comparisons. We conclude our results in Sec.~\ref{conclusion}.

\section{Generally covariant geometric momentum and geometric potential on a curved surface}\label{geometric}

Specifically, we consider a Dirac fermion constrained to move on a curved surface with a formal parametric equation ${\bf{r}}(\varphi,\theta)=(x,y,z)$, where $x,y,z$ can be the functions of either $\theta$ or $\varphi$. According to the definiton of generally covariant geometric momentum, we first calculate the natural basis of the curved surface by ${\bf{r}}_{\theta}=\frac{\partial {\bf{r}}}{\partial \theta}$ and ${\bf{r}}_{\varphi}=\frac{\partial {\bf{r}}}{\partial \varphi}$, and the unit normal vector is expressed as $
{\bf{n}}=\frac{{\bf{r}}_{\theta} \times {\bf{r}}_{\varphi}}{|{\bf{r}}_{\theta} \times {\bf{r}}_{\varphi}|}$, which leads to the metric $g_{\mu\nu}=\bf{r}_{\mu} \cdot \bf{r}_{\nu}$. One can subsequently obtain the inverse component of the natural frame ${\bf{r}}^{\mu}={\bf{r}}_{\mu}g^{\mu\nu}$, and the fundamental form of the curved surface is
 ${\bf{I}} =g_{\theta\theta}d{\theta}^2+g_{\varphi\varphi}d{\varphi}^2=(e^{1})^{2}+(e^{2})^{2}$, with $e^{1}$ and $e^{2}$ being the relative components of the dreibeins, which results in the transfer matrix $e_{\mu}^{a}$ between the space orthogonal coordinates and the local coordinates of the curved surface. 
% The non-zero terms in the spin connection $\omega_{\varphi}^{ab}$ are
% \begin{eqnarray}
% \omega_{\theta}^{12} &=& -e_{2}^{\nu}\left( \frac{\partial}{\partial \theta}e_{\nu}^{1}-\Gamma_{\theta\nu}^{\lambda}e_{\lambda}^{1}\right), %\nonumber \\
% % &=& -e_{2}^{\theta}\left[ \frac{\partial}{\partial \theta}e_{\theta}^{1}-\left(\Gamma_{\theta\theta}^{\theta}e_{\theta}^{1}+\Gamma_{\theta\theta}^{\varphi}e_{\varphi}^{1} \right)\right]-e_{2}^{\varphi}\left[\frac{\partial}{\partial\theta}e_{\varphi}^{1}-\left(\Gamma_{\theta\varphi}^{\theta}e_{\theta}^{1}+\Gamma_{\theta\varphi}^{\varphi}e_{\varphi}^{1} \right)\right],
% \end{eqnarray}
% where 
% \begin{eqnarray}
% \Gamma_{\theta\varphi}^{\theta}
% &=&\frac{1}{2}g^{\theta\xi}\left(\frac{\partial}{\partial\varphi}g_{\theta\xi}+\frac{\partial}{\partial \theta}g_{\xi\varphi}-\frac{\partial}{\partial\xi}g_{\theta\varphi} \right). 
%\nonumber\\
% &=& \frac{1}{2}g^{\theta\theta}\left( \frac{\partial}{\partial\varphi}g_{\theta\theta}+\frac{\partial}{\partial\theta}g_{\theta\varphi}-\frac{\partial}{\partial\theta}g_{\theta\varphi}\right)
% +\frac{1}{2}g^{\theta\varphi}\left( \frac{\partial}{\partial\varphi}g_{\theta\varphi}+\frac{\partial}{\partial\theta}g_{\varphi\varphi}-\frac{\partial}{\partial\varphi}g_{\theta\varphi}\right),
% \end{eqnarray}
% such that $\omega_{\theta}^{12} = -\omega_{\theta}^{21}= e_{2}^{\varphi}\Gamma_{\theta\varphi}^{\theta}e_{\theta}^{1}$. 
The covariant derivative and the gradient operator on a curved surface are $D_{\mu}=\partial_{\mu}+i\Omega_{\mu}$ and $\bf{r}^{\mu}\partial_{\mu}$, respectively, and the gauge part can be denoted by ${\bf{r}}^{\mu}\Omega_{\mu}=\left(x^{\prime},y^{\prime},z^{\prime}\right)$. Thus, we obtain the individual components of the generally covarient geometric momentum as
\begin{eqnarray}
p_{x} &=& \prod \nolimits_{x}-\hbar x^{\prime},\\
p_{y} &=& \prod\nolimits_{y}-\hbar y^{\prime},\\
p_{z} &=& \prod\nolimits_{z}-\hbar z^{\prime},
\end{eqnarray}
where $\prod\nolimits_{i}$ is the geometric momentum of the particle without spin.

For a fermion on a two-dimensional surface, the covariant Dirac equation can be generally written as
\begin{eqnarray}
-i\hbar\gamma^{\mu}D_{\mu}\Psi-\gamma^{0}m\Psi=0,
\end{eqnarray}
with $m \equiv m_{0} c$ denoting the reduced mass of the particle and $\gamma^{\mu}=e^{\mu a}\gamma_{a}$ representing the contravariant component of Dirac spin matrix under local coordinate, and the Hamiltonian is $H = -i\hbar\gamma^{\mu}D_{\mu}-\gamma^{0}m$. We assume that the geometric potential takes the form of a general $2 \times 2$ matrix
\begin{eqnarray}
V_{G}=a_{0}I+a_{x}\sigma_{x}+a_{y}\sigma_{y}+a_{z}\sigma_{z},
\end{eqnarray}
where $a_{0},a_{x},a_{y},a_{z}$ are functions of $\theta$ and $\varphi$, and hence the Hamiltonian including the geometric potential is
\begin{eqnarray}
H^\prime=H+V_{G}.
\end{eqnarray}
The existence of a geometric potential in a relativistic Hamiltonian can be examined by calculating the commutations of the three components, respectively, with
\begin{eqnarray}
[p_{i},H^\prime]=[p_{i},H]+[p_{i},V_{G}].    
\end{eqnarray}
Finally, we impose the quantization conditions
\begin{eqnarray}
\left({\bf{n}}\wedge[{\bf{p}},H^{\prime}]-[{\bf{p}},H^{\prime}]\wedge{\bf{n}}\right)=0,
\end{eqnarray}
according to which one can determine $(a_{0},a_{x},a_{y},a_{z})$ that the dynamical quantization conditions are met. We should, at this point, reach the explicit results for a Dirac fermion constrained on a two-dimensional hypersurface.

\section{A Dirac fermion on a pseudosphere and a helical surface} \label{example}

\begin{figure}[htbp]
    \includegraphics[width=0.25
\columnwidth]{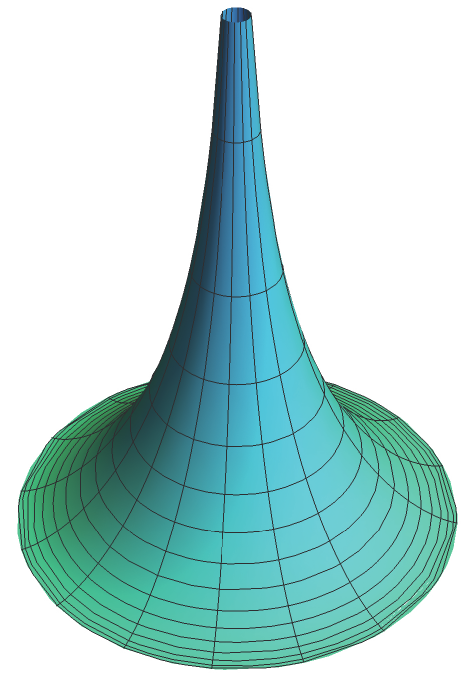}
    \caption{A pseudosphere with parametric equation ${\bf{r}}(u,v)=\left(\alpha\cos{u}\cos{v},\alpha\cos{u}\sin{v},\alpha[{\rm{ln}}(\sec{u}+\tan{v})-\sin{u}]\right)$, where $u \in [0,\frac{\pi}{2})$, $v\in [0, 2\pi)$, and $\alpha=1$.}
    \label{pseudosphere}
\end{figure}

We are now in position to consider two representative cases based on the above formalism, and verify the effectiveness of the enlarged canonical quantization scheme for constrained systems in curved surfaces. The parametric equation of a two-dimensional pseudosphere can be expressed as
\begin{eqnarray}
{\bf{r}}(u,v)=\left(\alpha\cos{u}\cos{v},\alpha\cos{u}\sin{v},\alpha[{\rm{ln}}(\sec{u}+\tan{u})-\sin{u}]\right),
\end{eqnarray}
where $u \in [0,\frac{\pi}{2})$, $v \in [0, 2\pi)$, and $\alpha$ is a constant, as depicted in Fig.~\ref{pseudosphere}. According to the natural basis and the relative components of the dreibeins, the nonzero term of the spin connection is
\begin{eqnarray}
\omega_{v}^{12}=-\omega_{v}^{21}=-\cos{u},
\end{eqnarray}
and the gauge parts of the gradient operator are
\begin{eqnarray}
{\bf{r}}^{u}\Omega_{u} &=& 0, \\
{\bf{r}}^{v}\Omega_{v} &=& \left(\frac{\sigma_{z}}{2\alpha}\sin{v},-\frac{\sigma_{z}}{2\alpha}\cos{v},0\right).
\end{eqnarray}
In addition, the mean curvature of the pseudosphere is given by $M=\frac{-\cot{u}+\tan{u}}{2\alpha}$, which in turn yields the respectively components of the generally covariant geometric momentum
\begin{eqnarray}
p_{x} &=& \prod\nolimits_{x}+\frac{\hbar}{2\alpha}\sigma_{z}\sin{v},  \\ \nonumber
p_{y} &=& \prod\nolimits_{y}-\frac{\hbar}{2\alpha}\sigma_{z}\cos{v},  \\ \nonumber
p_{z} &=& \prod\nolimits_{z},
\end{eqnarray}
where
\begin{eqnarray}
\prod\nolimits_{x} &=& -\frac{i\hbar}{\alpha}\left(-\frac{-\cot{u}+\tan{u}}{2}\cos{v}\sin{u}-\frac{\sin{v}}{\cos{u}}\partial_{v}-\cos{v}\frac{\cos{u}}{\tan{u}}\partial_{u}\right), \\
\prod\nolimits_{y} &=& -\frac{i\hbar}{\alpha}\left(-\frac{-\cot{u}+\tan{u}}{2}\sin{v}\sin{u}+\frac{\cos{v}}{\cos{u}}\partial_{v}-\sin{v}\frac{\cos{u}}{\tan{u}}\partial_{u}\right), \\
\prod\nolimits_{z} &=& -\frac{i\hbar}{\alpha}\left(-\frac{-\cot{u}+\tan{u}}{2}\cos{u}+\cos{u}\partial_{u} \right).
\end{eqnarray}

Accordingly, the transfer matrix is
\begin{eqnarray}
e^{\mu a} &=& e_{\nu}^{a}g^{\mu\nu} 
= \left(
\begin{array}{cc}
  \frac{1}{\alpha\cos{u}}   &  0\\
   0  & \frac{1}{\alpha\tan{u}}
\end{array}   \right),
% \left( 
% \begin{array}{cc}
%  \alpha\cos{u}    & 0 \\
%   0   & \alpha\tan{u}
% \end{array}
% \right) \left(
% \begin{array}{cc}
%   \frac{1}{\alpha^{2}\cos^{2}u}   &  0\\
%     0 & \frac{1}{\alpha^{2}\tan^{2}u}
% \end{array}
% \right) \nonumber\\
% &=& \left(
% \begin{array}{cc}
%   \frac{1}{\alpha\cos{u}}   &  0\\
%    0  & \frac{1}{\alpha\tan{u}}
% \end{array}   \right),
\end{eqnarray}
which leads to the contravariant component of Dirac spin matrix under local coordinate
\begin{eqnarray}
\gamma^{\mu} =
e^{\mu a}\gamma_{a} 
% = \left(
% \begin{array}{cc}
%   \frac{1}{\alpha\cos{u}}  & 0 \\
%    0  & \frac{1}{\alpha\tan{u}}
% \end{array} \right) \left(
% \begin{array}{c}
%   \sigma_{x}   \\ \sigma_{y} 
% \end{array}
%   \right) 
=
\left(
 \begin{array}{c}
 \frac{\sigma_{x}}{\alpha\cos{u}} \\ \frac{\sigma_{y}}{\alpha\tan{u}}
\end{array} \right),
\end{eqnarray}
and the Hamiltonian can be rewritten as
\begin{eqnarray}
H &=& -i\hbar\gamma^{\mu}\left(\partial_{\mu}+i\Omega_{\mu} \right)-\gamma^{0}m \nonumber\\
&=& -i\hbar\left[ \frac{\sigma_{x}}{\alpha\cos{u}}\left(\partial_{v}-\frac{i}{2}\sigma_{z}\cos{u} \right)+\frac{\sigma_{y}}{\alpha\tan{u}}\partial_{u}\right]+\sigma_{z}m \nonumber\\
&=& -\frac{i\hbar}{\alpha}\left[\frac{\sigma_{x}}{\cos{u}}\partial_{v}+\sigma_{y}\left(\frac{1}{\tan{u}}\partial_{u}-\frac{1}{2}\right)\right]+\sigma_{z}m.
\end{eqnarray}
After performing some computations, we obtain the following three equations for the geometric potential
\begin{eqnarray}
\frac{2i\hbar}{\alpha}\left[\frac{\sin{v}}{\tan{u}}\partial_{u}V_{G}-\cos{v}\partial_{v}V_{G}-\frac{1}{2}\cos{u}\cos{v}\left(a_{x}\sigma_{y}-a_{y}\sigma_{x}\right)\right] &=& 0, \\
\frac{2i\hbar}{\alpha}\left[\frac{\cos{v}}{\tan{u}}\partial_{u}V_{G}+\sin{v}\partial_{v}V_{G}+\frac{1}{2}\cos{u}\sin{v}\left(a_{x}\sigma_{y}-a_{y}\sigma_{x}\right)\right] &=& 0, \\
\frac{2i\hbar}{\alpha}\left[\tan{u}\partial_{v}V_{G}+\frac{1}{2}\sin{u}\left(a_{x}\sigma_{y}-a_{y}\sigma_{x}\right)\right] &=& 0.
\end{eqnarray}
It is obvious that the dynamic quantization condition is satisfied only when $(a_{0},a_{x},a_{y},a_{z})=(C_{1},0,0,C_{2})$, where $C_{1}$ and $C_{2}$ are constants. Thus, the geometric potential for a Dirac fermion constrained on a pseudosphere is
\begin{eqnarray}
V_{G}=C_{1}I+C_{2}\sigma_{z},
\end{eqnarray}
which is a constant matrix composed of the Pauli matrix in the $z$-direction and the identity matrix, and it might be related to the fact that the surface has no spin-orbit coupling or anisotropic curvature and the quantum nature of constraints \cite{maraner}. As shown in Ref.~\cite{yang} for two-dimensional curved surfaces of revolution, $C_{1}$ and $C_{2}$ can be set to zero by an appropriate energy reference shift as adding a constant
to the potential does not affect physical observables, which implies the absence of the geometric potential, i.e., $V_{G}=0$, and confirms the consistency of the Hamiltonian expression.

\begin{figure}[htbp]
    \includegraphics[width=0.25\columnwidth]{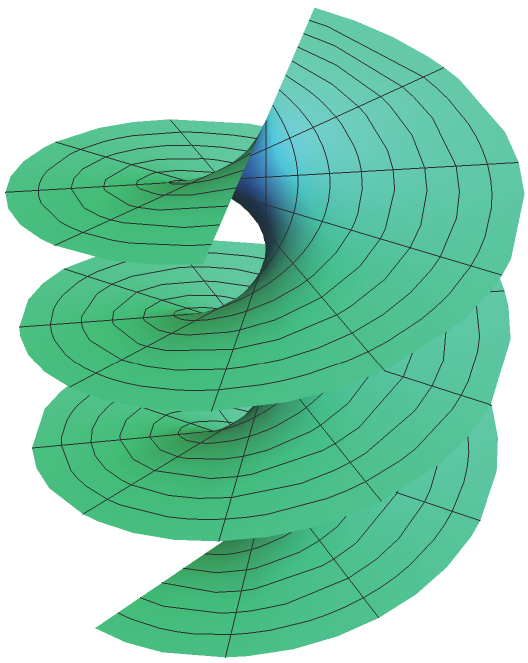}
    \caption{A helical surface with parametric equation ${\bf{r}}\left(u,v\right)=(u\cos{v},u\sin{v},\beta v)$, where $u \in (-\infty,\infty)$, $v \in (-\infty,\infty)$ and $\beta=1$.}
    \label{helical}
\end{figure}

With respect to the other example, the parametric equation of a helical surface in three-dimensional flat space under cartesian coordinate is given by
\begin{eqnarray}
{\bf{r}}\left(u,v\right)=(u\cos{v},u\sin{v},\beta v)
\end{eqnarray}
with $u \in (-\infty,\infty)$, $v \in (-\infty,\infty)$, and $\beta>0$, as sketched in Fig~\ref{helical}. The mean curvature is $M=0$, and by following the same lines as for the pseudosphere, one can obtain the explicit expressions for the generally covariant geometric momentum of the helical surface
\begin{eqnarray}
 p_{x} &=& \prod\nolimits_{x}+\frac{\hbar}{2}\sigma_{z}\frac{u^2\sin{v}}{\left(u^2+\beta^2\right)^{3/2}},  \\ \nonumber 
 p_{y} &=& \prod\nolimits_{y}-\frac{\hbar}{2}\sigma_{z}\frac{u^2\cos{v}}{\left(u^2+\beta^2\right)^{3/2}},  \\
 \nonumber
 p_{z} &=& \prod\nolimits_{z}-\frac{\hbar}{2}\sigma_{z}\frac{u\beta}{\left(u^2+\beta^2\right)^{3/2}},
\end{eqnarray}
with
\begin{eqnarray}
\prod\nolimits_{x} &=&-i\hbar\left(\cos{v}\partial_{u}-\frac{u\sin{v}}{u^2+\beta^2}\partial_{v}\right),\\
\prod\nolimits_{y} &=&-i\hbar\left(\sin{v}\partial_{u}+\frac{u\cos{v}}{u^2+\beta^2}\partial_{v}\right),\\
\prod\nolimits_{z} &=&-i\hbar\frac{\beta}{u^2+\beta^2}\partial_{v}.
\end{eqnarray}

Similarly, the transfer matrix can be obtained
\begin{eqnarray}
e^{\mu a} &=& e_{\nu}^{a}g^{\mu \nu} 
=\left(\begin{array}{cc}
   1  &  0\\
   0  & \frac{1}{\sqrt{u^{2}+\beta^{2}}}
\end{array}\right),
% \left(\begin{array}{cc}
%    1  & 0 \\
%    0  & \sqrt{u^{2}+\beta^{2}}
% \end{array}\right)
% \left(\begin{array}{cc}
%   1   & 0 \\
%    0  & \frac{1}{u^{2}+\beta^{2}}
% \end{array}\right) \nonumber\\
% &=&\left(\begin{array}{cc}
%    1  &  0\\
%    0  & \frac{1}{\sqrt{u^{2}+\beta^{2}}}
% \end{array}\right),
\end{eqnarray}
and the contravariant component of Dirac spin matrix under the local coordinate is
\begin{eqnarray}
\gamma^{\mu}=e^{\mu a}\gamma_{a}
% =\left(\begin{array}{cc}
%    1  & 0 \\
%    0  & \frac{1}{\sqrt{u^{2}+\beta^{2}}}
% \end{array}\right)
% \left(\begin{array}{c}
%      \sigma_{x} \\ \sigma_{y}
% \end{array}\right) 
=
\left(\begin{array}{c}
\sigma_{x} \\ \frac{\sigma_{y}}{\sqrt{u^{2}+\beta^{2}}}
\end{array}\right),
\end{eqnarray}
which results in the rewritten Hamiltonian
\begin{eqnarray}
H &=& -i\hbar\gamma^{\mu}\left(\partial_{\mu}+i\Omega_{\mu}\right)-\gamma^{0}m \nonumber\\
&=&-i\hbar\left[\sigma_{x}\partial_{u}+\frac{\sigma_{y}}{\sqrt{u^{2}+\beta^{2}}}\left(\partial_{v}-\frac{i\sigma_{z}}{2}\frac{u}{\sqrt{u^{2}+\beta^{2}}}\right)\right]+\sigma_{z}m \nonumber\\
&=&-i\hbar\left[\sigma_{x}\left(\partial_{u}+\frac{u}{2\left(u^{2}+\beta^{2}\right)}\right)+\frac{\sigma_{y}}{\sqrt{u^{2}+\beta^{2}}}\partial_{v}\right]+\sigma_{z}m.
\end{eqnarray}
The corresponding components of the geometric potential are thus straightforward
\begin{eqnarray}
2i\hbar\left[\frac{u\sin{v}}{\sqrt{u^2+\beta^2}}\partial_{u}V_{G}+\frac{\cos{v}}{\sqrt{u^2+\beta^2}}\partial_{v}V_{G}+\frac{u\cos{v}}{u^2+\beta^2}\left(a_{x}\sigma_{y}-a_{y}\sigma_{x}\right)\right] &=& 0, \\
2i\hbar\left[-\frac{u\cos{v}}{\sqrt{u^2+\beta^2}}\partial_{u}V_{G}+\frac{\sin{v}}{\sqrt{u^2+\beta^2}}\partial_{v}V_{G}+\frac{u\sin{v}}{u^2+\beta^2}\left(a_{x}\sigma_{y}-a_{y}\sigma_{x}\right)\right] &=& 0, \\
-2i\hbar\frac{\beta}{\sqrt{u^2+\beta^2}}\partial_{u}V_{G} &=& 0.
\end{eqnarray}
The general solutions of the above equations are $(a_{0},a_{x},a_{y},a_{z})=(C_{3},0,0,C_{4})$, where $C_{3}$ and $C_{4}$ are constant. Similarly, we obtain the geometric potential for a Dirac fermion confined on a helical surface
\begin{eqnarray}
V_{G}=C_{3}I+C_{4}\sigma_{z},
\end{eqnarray}
which also indicates no existence of the geometric potential. While the helical surface is not a curved surface of revolution, the result coincides with that of the pseudosphere. For other hypersurfaces, the characteristics may differ, which potentially offers insights into practical explorations.

\section{Conclusions and Discussions}\label{conclusion}

The fundamental and dynamical quantization conditions based on the quantization scheme of the classical system are applicable to a relativistic particle constrained to move on a curved surface. Within this framework, we have obtained the generally covariant geometric momentum of a Dirac fermion confined to a two-dimensional hypersurface, and demonstrated that no geoemtric potential exists in the cases of both the pseudosphere and the helical surface. The resulting geometric potentials are constant matrices independent of the parameters, and can be composed of the Pauli matrix in the $z$-direction and the identity matrix.

The non-existence of a geometric potential in these examples is probably related to the conformal flatness of two-dimensional surfaces, which eliminates curvature-induced geometric effects. This scenario might also arise in other curved systems where the gauge field or vector potential is constrained to be constant, for example, a uniform magnetic field in the $z$-direction and a system with no spin-orbit coupling or other geometric phases. Although a general result for Dirac fermions constrained on arbitrary two-dimensional hypersurfaces have yet to be established, it is convenient to resolve the curved systems with explicit parametric equations based on the proposed formalism. The related properties for other spin particles constrained on curved geometries remain interesting open questions, and are worthy of further investigations.

\section*{Acknowledgments}
We thank Professor Quan-Hui Liu for the insightful discussions. This work was supported by the Natural Science Foundation of Sichuan Province (Grant No. 2023NSFSC1330) and the Natural Science Research Start-up Foundation of Recruiting Talents of Nanjing University of Posts and Telecommunications (Grant No. NY223065).

% \begin{appendix}

% \section{}

% \end{appendix}

\end{document}